\documentclass[pre]{revtex4}
\usepackage[dvips]{graphicx}
\begin{document}

\title{Scaling Relations of Viscous Fingers in Anisotropic Hele-Shaw Cells}

\author{Hidetsugu Sakaguchi and Kazutaka Noto}

\affiliation{Department of Applied Science for Electronics and Materials, \\
Interdisciplinary Graduate School of Engineering Sciences, Kyushu
University, \\
Kasuga, Fukuoka 816-8580}
\begin{abstract}
Viscous fingers in a channel with surface tension anisotropy are numerically studied. Scaling relations between the tip velocity $v$, the tip radius $\rho$ and the pressure gradient $P_x$ are investigated for two kinds of boundary conditions of pressure, when $v$ is sufficiently large. The power-law relations for the anisotropic viscous fingers are compared with two-dimensional dendritic growth. The exponents of the power-law relations are theoretically evaluated.
\end{abstract}
\maketitle
\section{Introduction}
 Growth patterns in Laplace fields and diffusion fields have been intensively studied.\cite{rf:1,rf:2} In the problem of crystal growth from melt, dendrites grow in a diffusion field of temperature.  A steadily growing dendrite is characterized by the radius of curvature $\rho$ in the tip region and the growth velocity $v$. In the case of the absence of surface tension, only the product of $\rho$ and $v$ is determined as $\rho v=2DP(\Delta)\propto \Delta^2$ in two dimensions, where $\Delta$ is the supercooling and $D$ is the diffusion constant.  The surface tension anisotropy determines the value of $\rho$ and $v$ uniquely.  The tip radius and the velocity satisfy the scaling relations $v\propto \Delta^4$ and $\rho\propto \Delta^{-2}$ in two dimensions. As a result, $v\rho^2$ is constant, even when $\Delta$ is changed.\cite{rf:3,rf:4} 

 On the other hand, viscous fingers grow in the Hele-Shaw cell between two plates, where the pressure satisfies the Laplace equation.\cite{rf:5}  
The pressure gradient is a control parameter in the problem of viscous fingering, which corresponds to the supercooling $\Delta$ in the problem of crystal growth. In a channel of finite width, Saffman and Taylor found a steadily growing solution of the form~\cite{rf:6}  
\begin{equation}
x=\frac{1-\lambda}{\pi}\ln\frac{1}{2}\left (1+\cos\frac{\pi y}{\lambda}\right ).\end{equation}
The effect of surface tension is not considered in the analysis performed by the Saffman and Taylor. In the case of the absence of surface tension, the finger width $\lambda$ cannot be determined uniquely. McLean and Saffman solved numerically steadily growing solutions of viscous fingers in the case of finite surface tension.\cite{rf:7} The finger width is uniquely selected, even if anisotropy is absent.  However, the finger width $\lambda$ approaches 0.5, when the tip velocity $v$ and pressure gradient are increased to infinity.    
 
 Dendritic patterns were observed even in experiments of viscous fingering in Hele-Shaw cells perturbed by etching one of their plates.\cite{rf:8,rf:9,rf:10,rf:11,rf:12} The etching induces some anisotropy. Matsushita and Yamada,~\cite{rf:11} and Couder~\cite{rf:12} found the relation $v\rho^2=$const. for a single viscous finger in a certain parameter range.  The scaling relation is the same as the dendritic growth in crystal growth. They considered that dendritic patterns  in both  crystal growth and viscous fingering have the same properties. 
Kessler et al. and Dorsey and Martin studied numerically anisotropic viscous fingering by using the solvability condition or solving a nonlinear eigenvalue problem.\cite{rf:13,rf:14} Kessler et al. predicted a scaling relation by WKB approximation as $\lambda\sim v^{1/4}$ in the limit of large $v$.\cite{rf:13}  The relations of curvature and velocity, or curvature and pressure gradient, were not reported in these simulations. 

In a previous study, we investigated some relations between $\rho$, $v$ and the pressure gradient $P_x$ for a range of relatively small $v$ \cite{rf:15} using the original method developed by McLean and Saffman.\cite{rf:7} 
In this study, we continue the numerical simulation and show scaling relations in the range of sufficiently large $v$, where some universal scaling relations are expected. The purpose of this study is to point out some difference between the dendritic growth in anisotropic viscous fingering and that in crystal growth from the scaling relations. We consider that the difference comes from the difference between the diffusion field and the Laplace field.

\section{Model Equation and Boundary Conditions}
The model equation and boundary conditions are the same as those reported in the previous paper.\cite{rf:15} We briefly explain them, and the details were written in refs.7 and 15. Viscous fingers with the constant pressure $p_0$ are pushed into a viscous fluid with the viscosity $\mu$ in a Hele-Shaw cell.  
The mean velocity ${\bf u}$ in the viscous fluid satisfies Darcy's law:
${\bf u}=-(b^2/12\mu)\nabla p=\nabla \phi$, where $p$ is the pressure, $\phi$ is the velocity potential, and $b$ is the spacing between the top and bottom plates of the Hele-Shaw cell.  From the incompressibility of the fluid, $\phi$ satisfies the Laplace equation.
A viscous finger is assumed to grow in a channel of width $2a$. 
Two kinds of boundary conditions for the pressure field $p$ at the interface of the fingers have been studied.
\begin{equation}
\Delta p=\frac{T(\hat{\theta})}{R}.
\end{equation}
\begin{equation}
\Delta p=\frac{\pi}{4}\frac{T(\hat{\theta})}{R}+\frac{2T(\hat{\theta})}{b}[1+3.8\{\mu v_n/T(\hat{\theta})\}^{2/3}],
\end{equation}
where $T(\hat{\theta})$ denotes the anisotropic surface tension, $v_n$ is the velocity normal to the surface, $\hat{\theta}$ denotes the angle of the tangent vector of the interface from the $x$-axis, and $R$ is the radius of curvature on the surface. ($\rho$ is the radius of curvature at the tip position.) The boundary condition (2) represents the Young-Laplace relation. 
The boundary condition (2) is equivalent to the Gibbs-Thomson effect in the problem of crystal growth, if the pressure is replaced with the temperature.   
Most authors, such as McLean and Saffman,\cite{rf:7} and Kessler et al.\cite{rf:13} used this type of boundary condition for theoretical and numerical analyses. 
This boundary condition is useful for comparing the dendritic growth between the diffusion field and the Laplace field. 
The Young-Laplace relation is exactly satisfied, when the two fluids are stationary. Later, Park and Homsy derived the boundary condition (3) for the moving interface, when the ratio of the viscosities of two fluids is $O(Ca^{1/3})$.\cite{rf:16} Here, $Ca$ is the capillary number $Ca=\mu v/T$.  This boundary condition is more realistic for the moving interface, however, rather complicated. Even this boundary condition is not available, when the capillary number (velocity) becomes too large. The boundary condition (2) is obtained, if the second term of eq.~(3) is neglected, and $(\pi/4)T$ in eq.~(3) is replaced with $T$.  We study viscous fingers using the two boundary conditions and discuss the difference of the scaling relations caused by the boundary conditions.

The dimensionless velocity vector $(\hat{U},\hat{V})$ is expressed using the complex number as $\hat{U}-i\hat{V}=\hat{q}e^{-i\hat{\theta}}$.  The arclength $\hat{S}$ from the tip position along the interface is changed into another variable $s$ satisfying $0\le s\le 1$. At the finger tip, $s=1$ and $s=0$ at $x=-\infty$ on the interface.    
Because $\log \hat{q}-i(\hat{\theta}-\pi)$ is an analytic function, Cauchy's integral theorem yields
\begin{equation}
\log (\hat{q}(s))=-\frac{1}{\pi}\int_0^1\frac{\hat{\theta}(s^{\prime})-\pi}{s^{\prime}-s}ds^{\prime}.
\end{equation}
As an anisotropy of the surface tension, the fourfold rotational symmetry of the form $T(\hat{\theta})=T_0(1-\epsilon\cos4\theta)$ is assumed, where $\theta=\hat{\theta}-\pi$. The boundary condition (2) for the pressure is expressed as
\begin{equation}
\kappa qs\frac{d}{ds}\left \{(1-\epsilon\cos4\theta)qs\left (\frac{d\theta}{ds}\right )\right\}-q=-\cos\theta,
\end{equation}
where $q=(1-\lambda)\hat{q}$, $\kappa=T_0b^2\pi^2/\{12\mu va^2(1-\lambda)^2\}$.  
The boundary condition (3) is expressed as
\[
\kappa qs\frac{d}{ds}\left \{(1-\epsilon\cos4\theta)qs\left (\frac{d\theta}{ds}\right )\right\}+\kappa(1-\lambda)g(-qs)\frac{d}{ds}\left[ (1-\epsilon\cos4\theta)\left \{1+3.8\left (\frac{\mu \pi v(-\sin\theta)}{4T_0(1-\epsilon\cos4\theta)}\right )^{2/3}\right \}\right ] \]
\begin{equation}
-q=-\cos\theta, 
\end{equation}
where $T(\hat{\theta})=(4/\pi)T_0(1-\epsilon\cos4\theta)$ and $\kappa=T_0b^2\pi^2/\{12\mu va^2(1-\lambda)^2\}$ are assumed for the coefficient of the first term in eqs.~(5) and (6) to take the same form. The parameter $g$ is defined as $g=8a/(\pi^2b)$. 
The solutions to the coupled equations of eqs.~(4) and (5) (or eqs.~(4) and (6)) are numerically solved with the modified Newton method by discretizing the interface with $N=300$ intervals. If $q$ and $\theta$ are obtained, the finger interface is expressed through the relation
\begin{equation}
\hat{x}+i\hat{y}=-\frac{1-\lambda}{\pi}\int_s^1\frac{e^{i\theta}}{sq}ds.
\end{equation}

\begin{figure}[htb]
\begin{center}
\includegraphics[height=4.5cm]{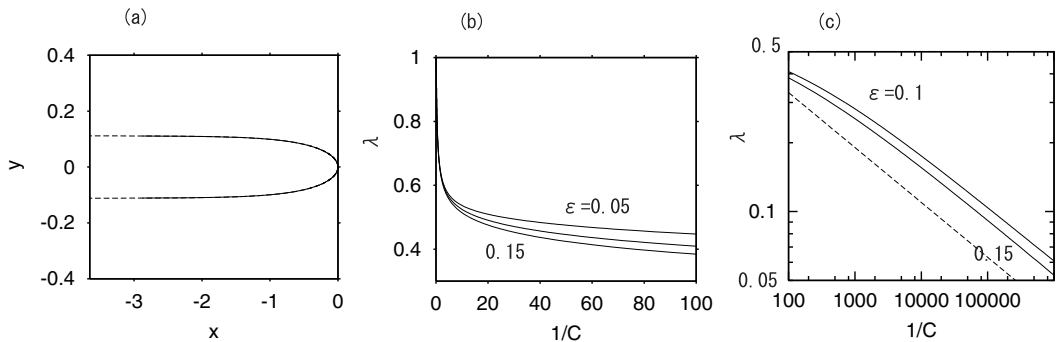}
\end{center}
\caption{(a) Viscous finger (solid curve) for $\epsilon=0.15$ and $1/C=42843$, and the Saffman-Taylor solution (dashed curve) with $\lambda=0.1115$. (b) Finger width $\lambda$ as a function of $1/C$ for $\epsilon=0.05,0.1$ and 0.15. (c) Double-logarithmic plot of $1/C$ vs $\lambda$  for $\epsilon=0.1$ and 0.15.}
\label{f1}
\end{figure}
\section{Scaling Relations for Anisotropic Viscous Fingers}
Figure 1(a) displays a viscous finger obtained numerically for $\epsilon=0.15$ and $1/C=12\mu va^2/(T_0b^2\pi^2)=42843$, which is a quantity proportional to the capillary number $Ca=\mu v/T$ and therefore the tip velocity $v$. The width of the finger is $\lambda=0.1115$.
 The dashed curve denotes the Saffman-Taylor solution with $\lambda=0.1115$. 
The numerically obtained solution is well approximated using the Saffman-Taylor solution when $\lambda$ is small. 
   Figure 1(b) displays $\lambda$ as a function of $1/C$ for $\epsilon=0.05,0,1$ and $0.15$. For $\epsilon=0$,  the finger width is always larger than 0.5. However, the finger width becomes smaller than 0.5 owing to the surface tension anisotropy.
Figure 1(c) displays a double-logarithmic plot of $1/C$ and $\lambda$ for $\epsilon=0.1$ and 0.15 in the range of $1/C>100$. The dashed curve denotes a power of $\lambda=(1/C)^{0.24}$. A power-law relation appears in the range of large $1/C$. The exponent 0.24 is very close to the exponent 0.25 predicted by Kessler et al. with WKB approximation. 
\begin{figure}[htb]
\begin{center}
\includegraphics[height=5.5cm]{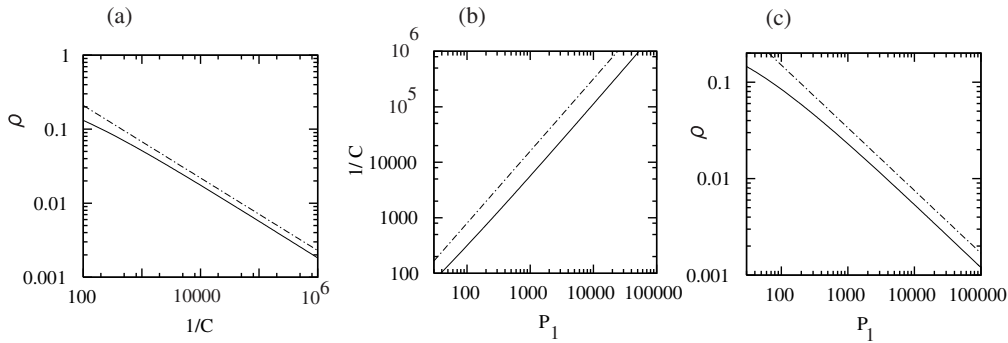}
\end{center}
\caption{(a) $1/C$ vs $\rho$. (b) $P_1$ vs $1/C$. (c) $P_1$ vs $\rho$ for $\epsilon=0.4$. The dashed lines in (a), (b), and (c) respectively denote power-law relations  with exponents -0.49, 1.3, and -0.65.}
\label{f2}
\end{figure}
Figures 2(a)-2(c) display double-logarithmic plots of (a) $1/C$ vs the tip radius $\rho$, (b) $P_1$ vs $1/C$, and (c)  $P_1$ vs $1/C$ for $\epsilon=0.15$, where $P_1=\lambda/C$.  The pressure gradient $P_x$ at $x\rightarrow \infty$  is a control parameter that corresponds to the supercooling $\Delta$ in the  problem of crystal growth. The pressure gradient at $x\rightarrow \infty$ is evaluated as $-\partial p/\partial x=(12\mu/b^2)\partial \phi/\partial x\sim (12\mu v\lambda)/b^2\propto \lambda/C$. On the other hand, the tip velocity $v$ is proportional to $1/C$. Figure 2 therefore shows the power-law relations $\rho\propto 1/v^{0.49}$, $v\propto P_x^{1.3}$, and $\rho\propto 1/P_x^{0.65}$. The exponent of $\rho$ vs $1/v$ is consistent with the scaling relation $\rho\propto 1/v^{0.5}$ for the dendrite  in the crystal growth. However, the other exponents are different from the scaling relation for the dendrite in the two-dimensional crystal growth: $v\propto \Delta^4$ and $\rho\propto 1/\Delta^2$. 
The difference of the scaling relations comes from the difference between the diffusion field and the Laplace field.  In the Laplace field, there is no characteristic length scale such as the diffusion length in the diffusion field. 
As a result, the pressure should decrease constantly with a constant pressure gradient $P_x$ in the range of $x\rightarrow +\infty$ in an infinitely long channel. 

We can evaluate the exponents of the scaling relations. As shown in Fig.~1(a), the Saffman-Taylor solution is a very good approximation for the anisotropic viscous finger with a large tip velocity $v$. Then the tip radius $\rho$ is proportional to $d^2x/dy^2$ at $y=0$ for eq.~(1), which is evaluated as $\lambda^2/\{\pi^2(1-\lambda)\}$. For a sufficiently small $\lambda$, $\rho\sim \lambda^2$. The prediction $\lambda\propto (1/v)^{1/4}$ by Kessler et al. yields the relation $\rho\propto 1/v^{0.5}$ or $v\rho^2=$const. The pressure gradient $-\partial p/\partial x$ is proportional to $\lambda/C\propto \lambda v$, which leads to the scaling relations: $P_x\propto \lambda v \propto v^{3/4}\propto \rho^{-3/2}$. That is, $\rho\propto v^{-1/2}, v\propto P_x^{4/3}$ and $\rho\propto P_x^{-2/3}$. These scaling relations are consistent with the numerical results shown in Figs.~1(c), and 2(a)-2(c) for $\epsilon=0.15$. 
The scaling relations other than $\lambda\propto (1/v)^{1/4}$ have not yet been reported. The exponents evaluated in the previous study~\cite{rf:15} were different from the present values, because they were evaluated when the tip velocity $v$ was not sufficiently large.

\begin{figure}[htb]
\begin{center}
\includegraphics[height=5.cm]{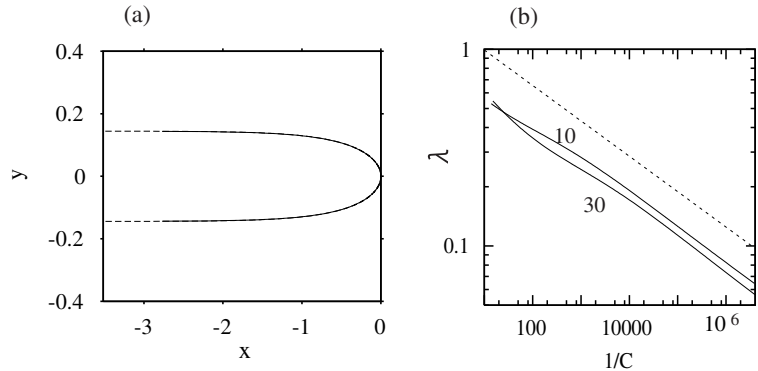}
\end{center}
\caption{(a) Viscous finger (solid curve) for $\epsilon=0.15, g=10$, and $1/C=47156$, and the Saffman-Taylor solution (dashed curve) with $\lambda=0.1445$. (b) Double-logarithmic plot of the finger width $\lambda$ as a function of $1/C$ for  $g=10$ and 30 at $\epsilon=0.15$. The dashed line denotes a power law with the exponent -0.18.} 
\label{f3}
\end{figure}
Next, we investigate the case of the boundary condition (3). 
Figure 3(a) displays a viscous finger for $\epsilon=0.15,g=10$, and $1/C=47156$. It is also well approximated using the Saffman-Taylor solution. 
Figure 3(b) displays a double-logarithmic plot of $\lambda$ as a function of $1/C=12\mu va^2/(T_0b^2\pi^2)$ at $\epsilon=0.15$ for $g=10$ and 30. 
We find the power-law relation $\lambda\propto 1/v^{\alpha}$ in the  asymptotic region of $1/C\rightarrow \infty$. The exponent $\alpha$ is estimated at 0.18, which is smaller than the exponent $\alpha=0.25$ for the case of the boundary condition (2).   
Figures 4(a)-4(c) display double-logarithmic plots of (a) $1/C$ vs $\rho$, (b) $P_1$ vs $1/C$  and (c) $P_1$ vs $\rho$ for $\epsilon=0.15$ and $g=10$. The dashed lines in Figs.~4(a)-4(c) denote the power-law relations  $\rho\propto 1/v^{0.36}$, $v\propto P_x^{1.21}$, and $\rho\propto 1/P_x^{0.44}$, respectively. 
For the boundary condition (3), dominant terms are 
\begin{equation}
\kappa(1-\lambda)g(-qs)3.8(1-\epsilon\cos4\theta)\frac{d}{ds}\left (\frac{\mu \pi v(-\sin\theta)}{4T_0(1-\epsilon\cos4\theta)}\right )^{2/3}=q-\cos\theta,
\end{equation}
in the limit of large $v$. Since $\kappa\propto 1/v$, the main term on the left-hand side is evaluated as $\kappa^{1/3}d\theta/ds$. The order estimate by WKB approximation yields $\lambda^2\sim \kappa^{1/3}$ or $\lambda\propto 1/v^{1/6}$. Because the Saffman-Taylor approximation is good for a large $v$, the tip radius $\rho\propto \lambda^2\propto 1/v^{1/3}$.  Then, the relation $v\rho^3=$const. is expected to be satisfied.  The exponents for the scaling relations $v\propto P_x^{\beta}$ and $\rho\propto 1/P_x^{\gamma}$ are expected to be  $\beta=6/5$ and $\gamma=2/5$, respectively. The numerically obtained exponents $\alpha=0.18$, $\beta=1.21$ and $\gamma=0.44$ are consistent with the order estimates $\alpha=0.167,\beta=1.2$, and $\gamma=0.4$. 
\begin{figure}[htb]
\begin{center}
\includegraphics[height=5.cm]{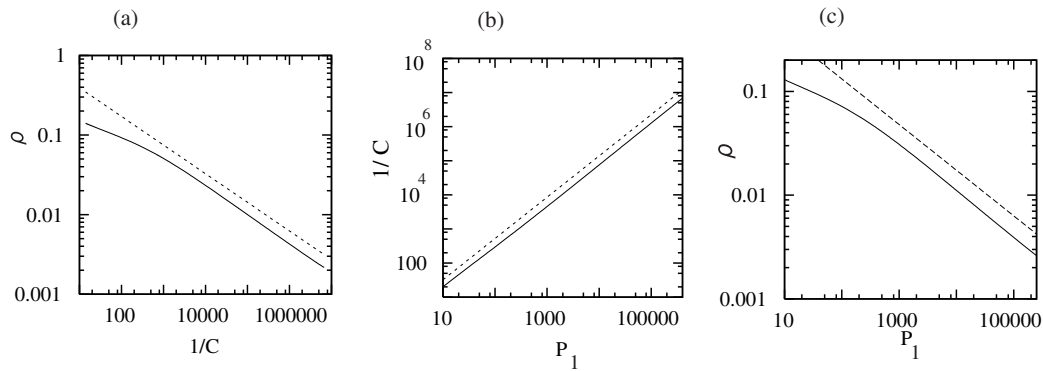}
\end{center}
\caption{Double-logarithmic plots of (a) $1/C$ vs $\rho$ for $g=10$ and $\epsilon=0.15$, (b) $P_1$ vs $1/C$ for $g=10$, and (c) $P_1$ vs $\rho$. The dashed lines in (a), (b), and (c) respectively denote power-law relations with exponents -0.36, 1.21, and -0.44.}
\label{f4}
\end{figure}

\section{Summary and Discussion}
We have numerically studied steadily growing viscous fingers with anisotropic surface tension in the range of large $1/C$ values.  We have confirmed the relation $v\rho^2=$const. even for anisotropic viscous fingering with the boundary condition (2); however, the exponents of the scaling relations $v\propto P_x^{\beta}$ and $\rho\propto 1/P_x^{\gamma}$ are respectively $\beta=4/3$ and $\gamma=2/3$  in the anisotropic viscous fingers, which are clearly different from the values $\beta=4$ and $\gamma=2$ for the dendrites in two-dimensional crystal growth. Because the boundary condition is the same, the difference of the scaling relations comes from the difference between the diffusion field and the Laplace field.  

If the boundary condition (3) is assumed, the relation $v\rho^2=$const. is not satisfied even in the limit of large $v$; however, $v\rho^{\delta}=$const. with $\delta\sim 1/(2\alpha)\sim 3$ is expected to be satisfied, because the velocity effect dominates in the boundary condition in the limit of large $v$. 

In the experiments of Matsushita and Yamada, and Couder, the relation of $v\rho^2$ appears in a relatively small range of $v$. They have also observed that the tip radius $\rho$ becomes nearly constant in the range of large $v$. Matsushita and Yamada suggested that the crossover originates from the second term including the $v$ of the boundary condition (3).  We have found the relation $v\rho^2=$const., but our numerical simulation is not always consistent with these experiments, in that we have not found $\rho=$const. in the range of large $v$ for the boundary condition (3). In the experiment of Matsushita and Yamada, a dendrite grows along a straight groove. In the experiment of Couder, a bubble is attached to the tip region of a viscous finger. We consider that the boundary conditions and experimental setting of these experiments are different from those in our numerical simulations. 

In the experiments by Ben-Jacob et al.~\cite{rf:8}, McCloud and Maher~\cite{rf:9}, and Honda et al.~\cite{rf:10}, many straight grooves are etched on the plates of the Hele-Shaw cell. Our numerical simulation corresponds to these experiments when $v$ is sufficiently small, where the surface tension dendrites appear. However, in these experiments, the surface tension dendrites disappear, instead, kinetic dendrites that grow along the direction of the groove  appear in the range of large $v$.  It is because the spacing $b$ in the groove regions is larger than that in the other regions, and therefore the mean velocity is larger owing to Darcy's law: ${\bf u}=-(b^2/12\mu)\nabla p$. Some kinetic effect of the large value of $v$ and the effect of the uniform spacing $b$ are considered in the boundary condition (3). However, the effect of the heterogeneity  of the spacing $b$ or the essential three-dimensional effect should be considered to discuss the kinetic dendrites growing along the direction of the grooves.  We are not sure whether the effect of the heterogeneity can be considered by the modification of the boundary condition through a $\theta$-dependent parameter, such as  $b(\theta)$, or direct three-dimensional simulations are necessary.

To summarize, we have numerically shown that the scaling relations change with the fields, in which patterns grow, and the boundary conditions. 
Our results are mathematical ones, and the relevance of the scaling relations to actual experiments remains insufficient now as discussed above. We hope that the scaling relations will be confirmed by suitable experiments

\end{document}